\documentclass[amsmath,amssymb,aps,prl,twocolumn,superscriptaddress,floatfix,sort&compress,longnamesfirst]{revtex4-2}
\usepackage{graphicx}
\usepackage{dcolumn}
\usepackage{bm}
\usepackage[dvipsnames]{xcolor}
\usepackage[colorlinks=true, linkcolor=MidnightBlue, urlcolor=MidnightBlue, citecolor=MidnightBlue]{hyperref}

\usepackage{setspace}
\begin{document}

\preprint{APS/123-QED}

\title{Suppression of Edge Localized Modes in ITER Baseline Scenario in EAST using Edge Localized Magnetic Perturbations}

\affiliation{Institute of Plasma Physics, Hefei Institutes of Physical Science, Chinese Academy of Sciences, Hefei 230031, China}
\affiliation{University of Science and Technology of China, Hefei 230026, China}
\affiliation{ITER Organization, Route de Vinon sur Verdon, 13115 St. Paul Lez Durance, France}
\affiliation{General Atomics, San Diego, California 92186, USA} 
\affiliation{Forschungszentrum J\"ulich, J\"ulich 52425, Germany}
\affiliation{Columbia University, New York, New York 10027, USA}
\affiliation{Donghua University, Shanghai 201620, China} 
\affiliation{Tsinghua University, Beijing 100084, China} 

\author{P. Xie}
\affiliation{Institute of Plasma Physics, Hefei Institutes of Physical Science, Chinese Academy of Sciences, Hefei 230031, China}
\affiliation{University of Science and Technology of China, Hefei 230026, China}
\affiliation{Forschungszentrum J\"ulich, J\"ulich 52425, Germany}
\author{Y. Sun}
\email{ywsun@ipp.ac.cn}
\affiliation{Institute of Plasma Physics, Hefei Institutes of Physical Science, Chinese Academy of Sciences, Hefei 230031, China}
\author{M. Jia}
\affiliation{Institute of Plasma Physics, Hefei Institutes of Physical Science, Chinese Academy of Sciences, Hefei 230031, China}
\author{A. Loarte}
\affiliation{ITER Organization, Route de Vinon sur Verdon, 13115 St. Paul Lez Durance, France}
\author{Y. Q. Liu}
\affiliation{General Atomics, San Diego, California 92186, USA}
\author{C. Ye}
\affiliation{Institute of Plasma Physics, Hefei Institutes of Physical Science, Chinese Academy of Sciences, Hefei 230031, China}
\author{S. Gu}
\affiliation{Institute of Plasma Physics, Hefei Institutes of Physical Science, Chinese Academy of Sciences, Hefei 230031, China}
\author{H. Sheng}
\affiliation{University of Science and Technology of China, Hefei 230026, China}
\affiliation{Institute of Plasma Physics, Hefei Institutes of Physical Science, Chinese Academy of Sciences, Hefei 230031, China}
\author{Y. Liang}
\affiliation{Forschungszentrum J\"ulich, J\"ulich 52425, Germany}
\author{Q. Ma}
\affiliation{Institute of Plasma Physics, Hefei Institutes of Physical Science, Chinese Academy of Sciences, Hefei 230031, China}
\affiliation{University of Science and Technology of China, Hefei 230026, China}
\author{H. Yang}
\affiliation{Institute of Plasma Physics, Hefei Institutes of Physical Science, Chinese Academy of Sciences, Hefei 230031, China}
\affiliation{University of Science and Technology of China, Hefei 230026, China}
\author{C. A. Paz-Soldan}
\affiliation{Columbia University, New York, New York 10027, USA}
\author{G. Deng}
\affiliation{Institute of Plasma Physics, Hefei Institutes of Physical Science, Chinese Academy of Sciences, Hefei 230031, China}
\affiliation{University of Science and Technology of China, Hefei 230026, China}
\author{S. Fu} 
\affiliation{Institute of Plasma Physics, Hefei Institutes of Physical Science, Chinese Academy of Sciences, Hefei 230031, China}
\affiliation{University of Science and Technology of China, Hefei 230026, China}
\author{G. Chen}
\affiliation{Donghua University, Shanghai 201620, China}
\author{K. He}
\affiliation{Institute of Plasma Physics, Hefei Institutes of Physical Science, Chinese Academy of Sciences, Hefei 230031, China}
\author{T. Jia}
\affiliation{Institute of Plasma Physics, Hefei Institutes of Physical Science, Chinese Academy of Sciences, Hefei 230031, China}
\author{D. Lu}
\affiliation{Institute of Plasma Physics, Hefei Institutes of Physical Science, Chinese Academy of Sciences, Hefei 230031, China}
\affiliation{University of Science and Technology of China, Hefei 230026, China}
\author{B. Lv}
\affiliation{Institute of Plasma Physics, Hefei Institutes of Physical Science, Chinese Academy of Sciences, Hefei 230031, China}
\author{J. Qian}
\affiliation{Institute of Plasma Physics, Hefei Institutes of Physical Science, Chinese Academy of Sciences, Hefei 230031, China}
\author{H.H. Wang}
\affiliation{Institute of Plasma Physics, Hefei Institutes of Physical Science, Chinese Academy of Sciences, Hefei 230031, China}
\author{S. Wang} 
\affiliation{University of Science and Technology of China, Hefei 230026, China}
\author{D. Weisberg}
\affiliation{General Atomics, San Diego, California 92186, USA}
\author{X. Wu}
\affiliation{Institute of Plasma Physics, Hefei Institutes of Physical Science, Chinese Academy of Sciences, Hefei 230031, China}
\affiliation{University of Science and Technology of China, Hefei 230026, China}
\author{W. Xu} 
\affiliation{Institute of Plasma Physics, Hefei Institutes of Physical Science, Chinese Academy of Sciences, Hefei 230031, China}
\author{X. Yan}
\affiliation{Institute of Plasma Physics, Hefei Institutes of Physical Science, Chinese Academy of Sciences, Hefei 230031, China}
\affiliation{Forschungszentrum J\"ulich, J\"ulich 52425, Germany}
\author{Y. Yu} 
\affiliation{Institute of Plasma Physics, Hefei Institutes of Physical Science, Chinese Academy of Sciences, Hefei 230031, China}
\author{Q. Zang}
\affiliation{Institute of Plasma Physics, Hefei Institutes of Physical Science, Chinese Academy of Sciences, Hefei 230031, China}
\author{L. Zeng}
\affiliation{Tsinghua University, Beijing 100084, China}
\author{T. Zhang}
\affiliation{Institute of Plasma Physics, Hefei Institutes of Physical Science, Chinese Academy of Sciences, Hefei 230031, China}
\author{C. Zhou}
\affiliation{University of Science and Technology of China, Hefei 230026, China}
\author{Z. Zhou}
\affiliation{Institute of Plasma Physics, Hefei Institutes of Physical Science, Chinese Academy of Sciences, Hefei 230031, China}
\affiliation{University of Science and Technology of China, Hefei 230026, China}
\author{B. Wan}
\affiliation{Institute of Plasma Physics, Hefei Institutes of Physical Science, Chinese Academy of Sciences, Hefei 230031, China}
\author{the EAST Team}
\date{\today}
\begin{abstract}
We report the suppression of Type-I Edge Localized Modes (ELMs) in the EAST tokamak under ITER baseline conditions using $n = 4$ Resonant Magnetic Perturbations (RMPs), while maintaining energy confinement. 
Achieving RMP-ELM suppression requires a normalized plasma beta ($\beta_N$) exceeding 1.8 in a target plasma with $q_{95}\approx3.1$ and tungsten divertors. 
Quasi-linear modeling shows high plasma beta enhances RMP-driven neoclassical toroidal viscosity torque, reducing field penetration thresholds. 
These findings demonstrate the feasibility and efficiency of high $n$ RMPs for ELM suppression in ITER. 
\end{abstract}
\maketitle


\textit{Introduction.-} 
Controlling edge instabilities while maintaining high plasma energy confinement is crucial for the success of fusion reactors.
Type-I Edge Localized Modes (ELMs), driven by high pressure and current gradients during H-mode confinement \cite{connor1998magnetohydrodynamic}, pose significant threats due to transient heat pulses and abrupt energy loss for a future fusion reactor, e.g. ITER \cite{hawryluk2009principal,loarte2003characteristics,loarte2014progress}.
Application of small amplitude non-axisymmetric Resonant Magnetic Perturbations (RMPs) is an effective control method widely investigated in many tokamaks \cite{evans2004suppression,liang2007active,Kirk2010NF,jeon2012suppression,sun2016nonlinear,suttrop2018experimental}.
However, RMPs often cause substantial density pump-out and energy confinement degradation \cite{liang2007active,sun2016nonlinear,suttrop2018experimental,sun2016edge,park2022overview}, which may make it hard for ITER to achieve its mission with high fusion gain $Q=10$ \cite{loarte2014progress,hawryluk2009principal}. 
In 2018, ELM suppression in low input torque plasmas using $n=4$ RMPs was successfully demonstrated for the first time in EAST in a narrow edge safety factor $q_{95}$ window around 3.65 \cite{sun2021first}.
Here $n$ is toroidal mode number.
Nevertheless, it remains challenging to achieve ELM suppression in the ITER $Q=10$ baseline scenario with $q_{95} \approx 3$ in low input torque plasmas.   
Here, we report the recent achievement in the EAST tokamak on Type-I ELMs suppression using $n=4$ RMPs under conditions matching the ITER baseline scenario without confinement drop.
We also discover a key 3D neoclassical transport effect at high plasma normalized beta ($\beta_N \gtrsim 1.8$) from edge localized high $n$ RMPs in achieving this goal, aided by quasi-linear modeling of RMP field penetration.

\textit{Experiment.-}
This experiment was conducted in EAST with $q_{95}$ around 3.1 and a normalized plasma beta $\beta_N$ of 1.8-2.0, which are all very close to the ITER $Q=10$ baseline operational scenario. 
EAST is a superconducting tokamak with a major radius $R_0=1.8~\mathrm{m}$ and a minor radius $a=0.45~\mathrm{m}$.
A flexible in-vessel RMP coil system with two arrays of ($2\times 8$) coils was installed on the low field side in EAST \cite{sun2016nonlinear}.
We employ $n=4$ RMPs because of its edge localized characteristics and many advantages such as no obvious drop in energy confinement, reducing core tungsten concentration, compatibility with pellet fuelling, and steady-state divertor power flux control using gas fuelling and neon impurity seeding, \textit{etc.}, demonstrated in the past \cite{jia2021integrated}. 
Since 2020, both the lower and upper divertors in EAST have been upgraded to ITER-like tungsten divertors. 

\begin{figure}[th]
\centering
\includegraphics[width=0.4\textwidth]{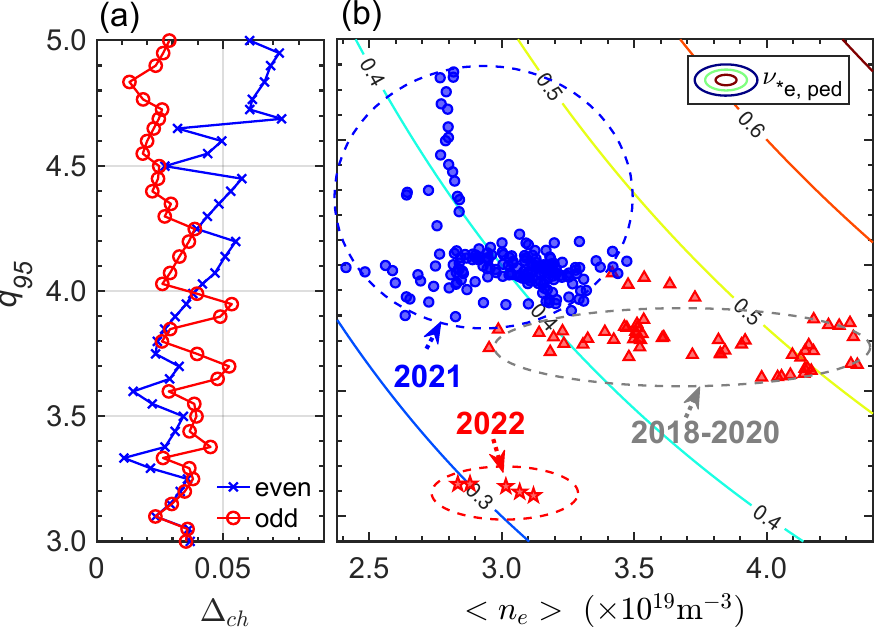}
\caption{\label{fig:window_modeling} (a) Linear modeling of ELM control windows, considering the plasma response to $n=4$ RMPs in EAST, showing the dependence of the edge stochastic layer width $\Delta_{ch}$ on $q_{95}$. (b) Experimental domain of ELM suppression discharges with $n=4$ RMPs in $q_{95}$ and line-integrated density.} 
\end{figure}

To guide the experimental identification of RMP-ELM suppression windows in EAST, linear modelling results on the $q_{95}$ dependence of plasma response to $n=4$ RMPs using the MARS-F code \cite{liu2010full,yang2016modelling,gu2019new,Xie_2023} have been performed before the experiments, as shown in Fig. \ref{fig:window_modeling}.  
Here, the normalized edge stochastic layer width  $\Delta_{ch}$, where the Chirikov parameter (indicating the overlap of neighboring resonant islands) is greater than 1, is used for evaluating perturbation strength at the pedestal \cite{sun2015modeling}. 
A reference equilibrium with $q_{95}=4.0$ is used to generate a series of equilibria with different $q_{95}$ but a fixed $\beta_N=$1.2.
There are multiple resonant windows, and the window also depends on coil parities, as shown in the figure. 
It is consistent with previous observations that ELM suppression at $q_{95} \approx 3.65$ was only achieved with odd parity $n=4$ RMPs \cite{sun2021first}, and a wide window ($q_{95}\geq4$) for ELM suppression with even coil parity \cite{Xie_2023}.
There  is also a peak of the plasma response near $q_{95} \approx 3$, though it is not as evident as for higher $q_{95}$.
Based on that, significant efforts have been made in recent experiments to extend ELM suppression into the $q_{95} \approx 3$ range towards ITER baseline requirements, including the optimization of high $\beta_N$ and low recycling conditions, and a major achievement was demonstrated recently using the odd parity $n=4$ RMPs.

\begin{figure}[h]
\centering
\includegraphics[width=0.4\textwidth]{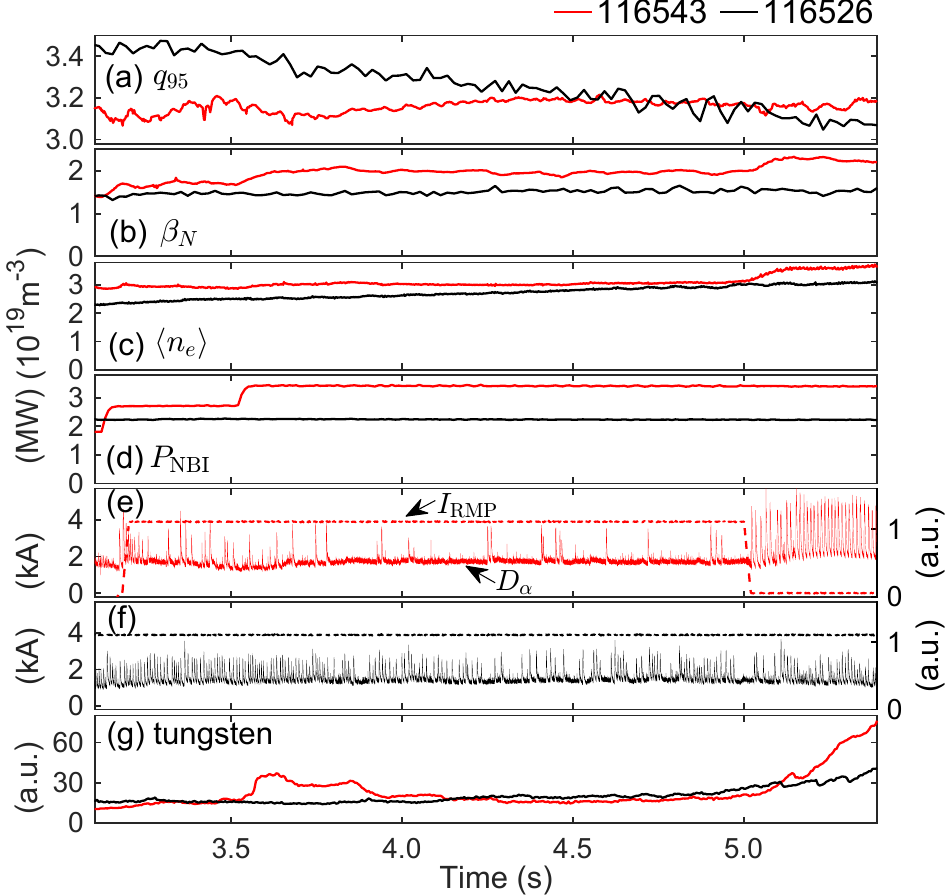}
\caption{\label{fig:shot116543} Temporal evolution of parameters in EAST discharges 116543 (red) and 116526 (black).
From top to bottom: (a) $q_{95}$, (b) $\beta_N$, (c) line-averaged plasma density, (d) neutral beam injection (NBI) injection power, (e) and (f)  $n=4$ coil current $I_\text{RMP}$ (dashed) and $D_\alpha$ (solid), and (g) tungsten concentration.}
\end{figure}

Suppression of Type-I ELMs using $n=4$ RMPs extrapolating favourably to the ITER baseline scenario has been achieved in EAST. 
One example of the normalized plasma beta effect on ELM suppression by $n = 4$ RMPs is shown in Fig.~\ref{fig:shot116543}. 
The toroidal magnetic field strength is $B_T \approx 1.5$ T, $q_{95} \approx 3.1$, and $\beta_N$ ranges from 1.7 to 2.0 during RMP application in discharge 116543.
The line-averaged plasma density is around $3.0\times 10^{19}~\mathrm{m}^{-3}$, corresponding to $40\%$ of the Greenwald density ($n_\text{GW}$), 
$H_{98}$ is $1.1$, energy confinement time is $\tau_E \approx 50~\mathrm{ms}$,
 and internal inductance is $l_i \approx 0.85$.
As shown in the figure, good ELM suppression has been achieved when $\beta_N$ exceeds around 1.8 at $t=3.8~\mathrm{s}$ after one additional neutral beam line switches on at $t = 3.5~\mathrm{s}$.
Although still some ELMs occasionally appeared, the suppression is sustained for a significantly longer duration than the energy confinement time.
As a comparison, discharge 116526 did not achieve ELM suppression with a lower $\beta_N \approx 1.5$, though a window with few ELMs around $t=4.5~\mathrm{s}$ suggests a marginal state for ELM suppression (For clearer illustration, the time axis has been slightly shifted).
This indicates that high plasma beta is essential for accessing ELM suppression via enhancing plasma response to RMPs. 
Tungsten concentration decreases during the application of RMPs and increases when distinct ELMs reappear after RMPs are turned off, indicating that ELM suppression effectively exhausts tungsten. 
ELM suppression has a small effect on energy confinement (see $\sim10\%$ $\beta_N$ increase after RMPs are turned off) and plasma density, as shown in Fig. 2.
In this case, the input torque is $1.8$ Nm, which is slightly higher than $1.1$ Nm, the extrapolated value for 33 MW of NBI in ITER. 
The normalized electron collisionality near the pedestal is around $\nu_\mathrm{*e,ped} \approx 0.3$, which is also a near-record low value for EAST. 
The observation of $n = 4$ RMP-ELM suppression at $q_{95} \approx 3.1$ in EAST, in which many parameters are very close to the values expected in the ITER baseline scenario, well validates the MARS-F prediction.

Experimental statistics at low $q_{95}=3.1$ show that ELM suppression is only achieved within the operational window of high $\beta_N$ ($\geq1.8$) and a low global particle recycling coefficient. 
In the higher beta cases, diagnostics show a clear increase in core plasma pressure, mainly due to higher temperature from increased NBI injection power.
Good wall conditioning is also found to be crucial for accessing ELM suppression, as hinted by the QH mode experiments in DIII-D \cite{grierson2014response}. 
To achieve ELM suppression, lithium powder has been applied in previous discharges to the ones in which ELM suppression is explored in order to reduce edge particle recycling. 
Changes in normalized density for different recycling conditions show that the density pedestal becomes wider at lower recycling rate cases, a tendency similar to ITER H-mode pedestals expected to have low $\nabla n_\text{ped}$.


\begin{figure}[ht]
\centering
\includegraphics[width=0.48\textwidth]{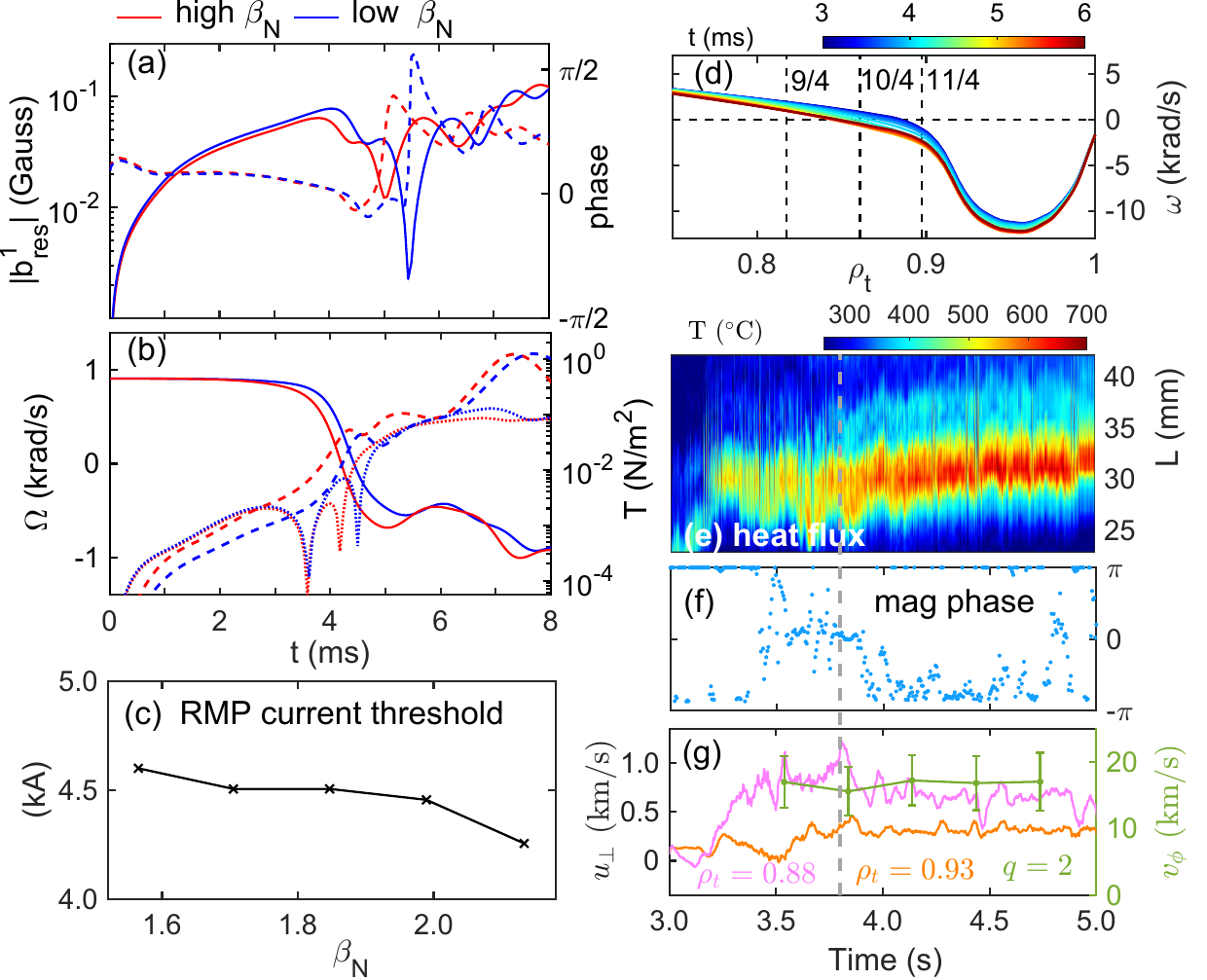}
\caption{\label{fig:qlinear} Quasi-linear modeling for plasma response. 
Time evolution: (a) amplitude (solid) and phase (dashed) of the resonant component at the 10/4 rational surfaces; (b) $\omega_E$ rotation (solid), NTV (dashed) and $j\times b$ (dotted) torque density at the surface. 
(c) Dependence of penetration threshold on $\beta_N$ from the modeling results.
(d) Evolution of the $\omega_E$ rotation profile. 
(e)-(g) Bifurcation phenomena in discharge 116543: divertor temperature distribution, magnetic field response to the external field, perpendicular flow near pedestal top, and toroidal rotation at $q=2$ surface.
The vertical dashed line indicates the transition time to ELM suppression.}
\end{figure}
\textit{Modeling.-} 
This work concentrates on the mechanisms of high $\beta_N$ in facilitating ELM suppression at low $q_{95}$.
Past experiments have shown that nonlinear bifurcations during the transition from ELM mitigation to suppression, such as changes in perpendicular rotation, magnetic signals, and electron pressure gradients \cite{sun2016nonlinear,paz2015observation,nazikian2015pedestal}, strongly indicate resonant field penetration.
This penetration is crucial for successful RMP application, directly influencing plasma edge conditions and pedestal stability, as supported by simulations and experimental observations \cite{hu2020wide,paz2019effect,hoelzl2021jorek}.  
Field penetration can be facilitated when the frequency of the perturbation field matches the mode frequency \cite{yu2008numerical}, and is influenced by plasma parameters like density and $\beta_N$ \cite{fitzpatrick1998bifurcated,ye2023effect}.
Previous studies show that there is no significant difference between the results in field penetration threshold obtained from the two-fluid model with $\omega_{e,\perp}$ and the single-fluid model with $\omega_E$ \cite{yu2008numerical,liu2014modelling,cole2006drift}.
We therefore employ the quasi-linear single-fluid model MARS-Q, utilizing experimental resistivity, to simulate the penetration process of external magnetic perturbations in the pedestal and investigate the $\beta_N$ effect. 
In MARS-Q model, the NTV torque couples with the MHD equations through the momentum transport equation \cite{liu2013toroidal}:
$$\frac{\partial \Delta L}{\partial t}=D(\Delta L)+T_\mathrm{NTV}+T_\mathrm{j\times b}+T_\mathrm{REY}, $$
where $L$ represents the toroidal angular momentum density, and $D(\Delta L)$ is the momentum diffusion term.
The analysis assumes that a momentum balance has been achieved before applying the RMP field.
The $\omega_E$ profile is chosen for the rotation model, and RMP current is set to ramp up to obtain the threshold for field penetration.

Plasma response modeling for different $\beta_N$ cases in this experiment is shown in Figure \ref{fig:qlinear}.
Figures \ref{fig:qlinear}(a) and (b) show the dynamics of resonant components ($m=nq$) of plasma response at the pedestal top for two cases, where the red one represents $\beta_N=2.1$, and the blue one represents $\beta_N=1.6$.
As the RMP current increases, the resonant components also increase initially linearly.
However, upon reaching a threshold, the amplitude of the resonant harmonics suddenly drops within approximately $1~\mathrm{ms}$, and the phase shifts (dashed lines). 
We define the moment of phase shift as the penetration time \cite{fitzpatrick1998bifurcated}.
It shows that the higher $\beta_N$ case triggers earlier field penetration, which means a lower threshold of RMP field amplitude.
Figure \ref{fig:qlinear}(b) shows the decreases in local $\omega_E$ rotation at the 10/4 surface, and the moment between phase shift and zero-rotation has a slight deviation. 
The rotation braking is mainly due to the Neoclassical Toroidal Viscosity (NTV) torque (dashed lines), with the smaller $j \times b$ torque (dotted lines) and the negligible Reynolds torque (no shown here).
This is consistent with previous works showing that higher $\beta_N$ enhances NTV torque at low collisionality \cite{sun2010neoclassical,yan2017neoclassical}. 
The modeling results are summarized in Fig. \ref{fig:qlinear}(c), showing that high $\beta_N$ reduces the RMP penetration threshold to values close to those applied in experiments.
Figure \ref{fig:qlinear}(d) shows the evolution of the $\omega_E$ profile for the high $\beta_N$ case during the field penetration, with rotation braking at the pedestal top ($10/4$ rational surface), reducing its amplitude and shear.


Figures \ref{fig:qlinear}(e)-(g) present experimental evidence of nonlinear bifurcation in discharge 116543. 
Splitting of heat flux on the divertor, located at $36~
\mathrm{mm}$, is clearly observed during ELM suppression, as shown in Fig. \ref{fig:qlinear}(e).
This, together with the phase shift in the $n=4$ component of plasma response extracted from magnetic probe signals and the perpendicular flow measured with the Doppler backscattering system \cite{zhou2013microwave} near the pedestal top, provides strong support that the nonlinear bifurcation of edge magnetic field topology is necessary for accessing ELM suppression \cite{sun2016nonlinear} and is consistent with the results of modeling penetration. 
Additionally, the core plasma rotation maintains sufficiently high at the $q = 2/1$ surface to avoid the penetration of lower order error fields ($n =1$) leading to increased plasma disruptivity \cite{park20183d}. %

High $\beta_N$ enhances the plasma response and the NTV torque, facilitating field penetration, might explain the reduced threshold for perturbation field at higher $\beta_N$, which is important for high $n$ RMPs ELM suppression.
The results also suggest that when properly managed \cite{li2019toroidal}, NTV torque could be a powerful tool for modulating plasma edge flow, controlling ELMs, and stabilizing the pedestal region, while at the same time preserving core rotaion and not increasing resistivity with $n=4$ RMPs.


\begin{figure}[ht]
\centering
\includegraphics[width=0.48\textwidth,clip]{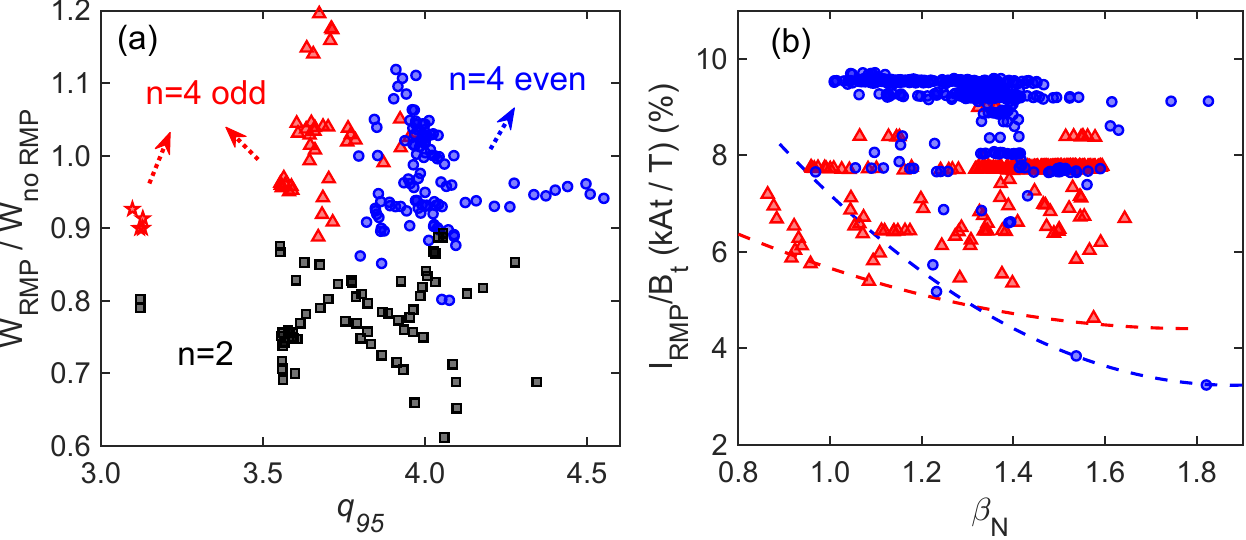}
	\caption{\label{fig:deltaW}(a) Stored energy change during ELM suppression with $n=2$ (black squares) and $n=4$ RMPs (other symbols) in EAST.
	(b) Current amplitude of $n=4$ RMPs normalized to $B_t$ required for entering ELM suppression in EAST at $q_{95}=4$ with even parity RMP (red), and $q_{95}=3.7$ with odd parity RMP (blue). 
	Dashed lines outline the boundaries of the RMP amplitude.} 
\end{figure}

\textit{Discussion.-} 
Integrating ELM suppression with good confinement is essential for future burning plasma devices.
Figure \ref{fig:deltaW}(a) shows that $n=4$ RMPs result in a much smaller drop in plasma stored energy during ELM suppression compared to $n = 2$ RMPs across different $q_{95}$ windows.
The energy loss during ELM suppression with $n=4$ RMPs is typically $\leq10\%$, while with $n=2$ RMPs is $20\%\sim30\%$.
The primary difference lies in the smaller reduction in density and temperature in the pedestal region for $n=4$ RMPs \cite{sun2016edge,Xie_2023}, which is possibly attributed to the more localized and modest plasma response to the high $n$ perturbation fields \cite{li2017toroidal,sun2021first}. 
Surprisingly, stored energy is even higher in some cases during ELM suppression with $n=4$ RMPs than in the ELMy H-mode phase, due to the formation of an ion Internal Transport Barrier (ITB) \cite{sun2023achievement}. 
The integration of ELM control and sustained high core confinement with ITB using high $n$ RMP is under investigation.

The normalized plasma beta effect is also observed in other RMP-ELM suppression windows in EAST.
Figure \ref{fig:deltaW}(b) presents the strength of $n=4$ RMPs when plasmas initially enter ELM suppression at different $\beta_N$, identified with AI assistance from over 200 discharges.
The ratio of the total turn current of RMP coils to the toroidal field evaluates the normalized field amplitude. 
The lower limit of RMP current is outlined with dashed lines, and it is shown that high $\beta_N$ lowers the RMP threshold for ELM suppression across various $q_{95}$ windows.
The RMP threshold for $q_{95}=3$ is the highest, with a normalized amplitude of $10\%$.
For ITER, a $60~\mathrm{kAt}$ level in the ELM control coils corresponds to $11\%$ in our normalization for the $Q =10$, $15~\mathrm{MA}$ scenario with $\beta_N\sim2$. 
Since the coils in ITER are designed for 90 kAt this implies that they should have sufficient capability to achieve ELM suppression. 
This finding is in agreement with MHD modelling for ITER \cite{becoulet2022non,hu2021nonlinear} showing robust ELM suppression in ITER for ELM control coil currents in excess of $50-60~\mathrm{kAt}$.
Considering ITER has three RMP coil arrays, while EAST has two, and all coils in ITER are powered independently, allowing fine-tuning of the relative phase, the EAST findings suggest ITER will have significant flexibility to demonstrate ELM suppression for $Q =10$ plasmas with $n = 3$ or 4.  
For other scenarios, such as $Q =5$ steady-state, this  flexibility is expected to be even greater on the basis of the EAST results because of the higher $\beta_N$ ($\approx 3$) \cite{polevoi2020reassessment}, leading to larger plasma response and higher $q_{95}\sim4.5$.  
This will allow some capability of the ELM control coils to be reserved for other control missions required in these high $\beta_N$ scenarios such as dynamic error field and resistive wall mode control, in addition to ELM control. 


In summary, remarkable results have been obtained in extending the $n=4$ RMP-ELM suppression window in EAST over the past four years, especially in suppressing Type-I ELMs under condition closely matching the ITER $15~\mathrm{MA}$ baseline scenario.
During ELM suppression, stored energy reduction is relatively insignificant ($\leq10\%$), and there was no tungsten accumulation.
High plasma beta is identified as key for achieving ELM suppression in this low $q_{95}$ experiment.
The high-$\beta_N$-enhanced NTV torque is revealed as the mechanism for braking the screening flow at the pedestal top, facilitating field penetration and ELM suppression.
These insights suggest an efficient approach for achieving edge stability in ITER and future devices with high plasma beta. 

\begin{acknowledgments}
This work is supported by the Natural Science Foundation of Anhui Province under Grant No. 2208085J39, the National Key R\&D Program of China under Grant No. 2017YFE0301100, the National Natural Science Foundation of China under Grant No. 12005261, No. 11875292, and No. 12105323 and the US DOE under DE-SC0021968.

The views and opinions expressed herein do not necessarily reflect those of the ITER Organization.
\end{acknowledgments}

\bibliography{References}

\end{document}